\documentclass[letterpaper, 10 pt, journal, twoside]{ieeetran}
\usepackage[T1]{fontenc}
\usepackage[latin9]{inputenc}
\usepackage{geometry}
\usepackage{color}
\usepackage{amsmath}
\usepackage{amsthm}
\usepackage{amssymb}
\usepackage[unicode=true,
 bookmarks=true,bookmarksnumbered=true,bookmarksopen=true,bookmarksopenlevel=1,
 breaklinks=false,pdfborder={0 0 0},pdfborderstyle={},backref=false,colorlinks=false]
 {hyperref}
\hypersetup{pdftitle={Your Title},
 pdfauthor={Your Name},
 pdfpagelayout=OneColumn, pdfnewwindow=true, pdfstartview=XYZ, plainpages=false}

\makeatletter
\theoremstyle{plain}
\newtheorem{thm}{\protect\theoremname}
\theoremstyle{definition}
\newtheorem{defn}[thm]{\protect\definitionname}
\theoremstyle{definition}
\newtheorem{example}[thm]{\protect\examplename}
\theoremstyle{plain}
\newtheorem{cor}[thm]{\protect\corollaryname}
\theoremstyle{plain}
\newtheorem{lem}[thm]{\protect\lemmaname}

\usepackage{epstopdf}

\usepackage[keeplastbox]{flushend}

\usepackage{tikz}
\usetikzlibrary{arrows}
\usetikzlibrary{shadows}

\usetikzlibrary{shapes}

\usetikzlibrary{arrows,decorations.pathmorphing}

\usepackage{hyperref}
\@ifundefined{rangeHsb}{\usepackage{xcolor}}{}
\usepackage{pdfsync}

\usepackage{kantlipsum,widetext}

\makeatother

\providecommand{\corollaryname}{Corollary}
\providecommand{\definitionname}{Definition}
\providecommand{\examplename}{Example}
\providecommand{\lemmaname}{Lemma}
\providecommand{\theoremname}{Theorem}

\begin{document}
\title{On the Controllability of Matrix-weighted Networks}
\author{Lulu Pan, Haibin~Shao,~\IEEEmembership{Member,~IEEE,} Mehran~Mesbahi,~\IEEEmembership{Fellow,~IEEE,}\\
Yugeng~Xi,~\IEEEmembership{Senior Member,~IEEE}, Dewei Li\thanks{This work is supported by the National Science Foundation of China
(Grant No. 61973214, 61590924, 61963030) and Natural Science Foundation
of Shanghai (Grant No. 19ZR1476200) and in part by the U.S. 
Air Force Office of Scientific Research (Grant No. FA9550-16-1-0022). (Corresponding
author: Haibin Shao)}\thanks{Lulu Pan, Haibin Shao, Yugeng Xi and Dewei Li are with the Department
of Automation and the Key Laboratory of System Control and Information
Processing Ministry of Education of China, Shanghai Jiao Tong University,
Shanghai 200240, China (\{llpan,shore,ygxi,dwli\}@sjtu.edu.cn). Mehran
Mesbahi is with the Department of Aeronautics and Astronautics, University
of Washington, Seattle, WA, 98195-2400, USA (mesbahi@uw.edu).}}
\maketitle
\begin{abstract}
This letter examines the controllability of consensus
dynamics on matrix-weighed networks from a graph-theoretic perspective.
Unlike the scalar-weighted networks, the rank of weight matrix introduces
additional intricacies into characterizing 
the dimension of controllable subspace for such networks. 
Specifically, we investigate
how the definiteness of weight matrices influences the dimension of
the controllable subspace. In this direction, graph-theoretic characterizations of
the lower and upper bounds on the dimension of the controllable subspace are provided
by employing,  respectively, distance partition and almost equitable partition of
matrix-weighted networks. Furthermore, the structure of an uncontrollable
input for such networks is examined. Examples are then provided to
demonstrate the theoretical results.
\end{abstract}

\section{Introduction}

Controllability of a dynamic system is a fundamental notion in control theory~\cite{kalman1963mathematical}. 
For multi-agent networks, controllability is closely related to the
graph-theoretic properties of the underlying network~\cite{tanner2004controllability,rahmani2009controllability,liu2011controllability}.
The controllability of multi-agent networks under nearest-neighbor
interactions has initially been examined in \cite{tanner2004controllability},
where it was shown that network connectivity can have adverse effects on controllability.
The influence of network symmetry of leader-following networks on
its controllability has been reported in \cite{rahmani2009controllability}.
\textbf{}Graph node partitions were subsequently employed to characterize the
upper bounds on the dimension of the controllable subspace of multi-agent networks
\cite{zhang2013upper,aguilar2017almost,egerstedt2012interacting};
analogous lower bounds have also been derived using distance partitions
\cite{zhang2013upper,yazicioglu2012tight}. Due to the difficulty in analyzing
controllability of general networks, controllability for 
special classes of networks has been an active area of research~\cite{aguilar2014graph,parlangeli2011reachability,ji2012leaders,chapman2014controllability,notarstefano2013controllability,hao2019controllability}.
Recently, controllability of multi-agent system on
signed networks (where cooperative and competitive interactions
coexist) has also received attention. For instance,
a graph-theoretic characterization of the upper bound on the dimension of the controllable
subspace for signed networks has been proposed using 
generalized equitable partition in \cite{sun2017controllability}.
In \cite{she2018topological}, sufficient conditions on the controllability
of signed path, cycle and tree networks have been derived. The controllability
problem on certain classes of signed networks is also studied in
\cite{guan2018controllability}; 
a comprehensive review on network controllability has been provided in~\cite{chen2019controllability}.

In the meantime, existing works on network controllability are mainly concerned with 
networks with scalar weighted edges; such network models are restrictive in characterizing
interdependence amongst subsets of the underlying node states~\cite{sun2018dimensional}.
Matrix-weighted networks are a natural extension of scalar-valued networks; they
have been examined in scenarios such
as graph effective resistance (motivated by distributed estimation and control)
\cite{barooah2006graph,tuna2017observability}, logical inter-dependency
of multiple topics in opinion evolution \cite{friedkin2016network,ye2018continuous},
bearing-based formation control \cite{zhao2015translational}, as well
as the array of coupled LC oscillators \cite{tuna2019synchronization}.
More recently, consensus and synchronization problems on matrix-weighted
networks have been examined in~\cite{trinh2018matrix,tuna2016synchronization,pan2018bipartite,su2019bipartite}.

Consensus protocol plays a vital role in cooperative
control of multi-agent networks, ensuring asymptotic alignment
on the states of the agents required for accomplishing
a global task via local interactions~\cite{mesbahi2010graph,olfati2004consensus,ren2005consensus,chung2018survey}.
In this letter, we examine the controllability of
multi-agent systems governed by consensus dynamics on matrix-weighted
networks. Although the matrix-weighted setup is a natural extension
of scalar-weighted networks, extending network controllability to the former case
is non-trivial. An essential distinction in this direction is that the rank of
the weighting matrix can range from zero up to its dimension.
In this note, we show how the definiteness of weight matrices
influences the dimension of the controllable  subspace for the corresponding network. 
Moreover, graph theoretic lower and upper bounds on the dimension of the controllable subspace of the
influenced consensus are provided--this is achieved by exploiting the distance partition and almost equitable
partition of matrix-weighted networks, extending results for scalar-weighted networks.

The remainder of this paper is organized as follows. The preliminary
notions used in the paper are introduced in \S 2. The problem formulation
is discussed in \S 3 followed by the characterization of
lower and upper bounds of the dimension of the controllable subspace
in \S 4 and \S 5, respectively. The structure of uncontrollable
input matrix is further discussed in \S 6. Examples are provided
in \S 7 followed by concluding remarks in \S 8.

\section{Preliminaries}

Let $\mathbb{R}$ and $\mathbb{N}$ be the set of real and natural
numbers, respectively. Denote $\underline{n}=\left\{ 1,2,\ldots,n\right\} $
for an $n\in\mathbb{N}$. A symmetric matrix $M\in\mathbb{R}^{n\times n}$
is positive definite, denoted by $M\succ0$, if $\boldsymbol{z}^{T}M\boldsymbol{z}>0$
for all nonzero $\boldsymbol{z}\in\mathbb{\mathbb{R}}^{n}$,
and is positive semi-definite, denoted by $M\succeq0$, if $\boldsymbol{z}^{T}M\boldsymbol{z}\ge0$
for all $\boldsymbol{z}\in\mathbb{\mathbb{R}}^{n}$.
The image and rank of a matrix $M$ are denoted by $\text{{\bf img}}(M)$
and $\text{{\bf rank}}(M)$, respectively. Denote by $\text{{\bf dim}}(\cdot)$
as the dimension of a vector space (or subspace) and $\text{{\bf diag}}\left\{ \cdot\right\} $
as the (block) diagonal matrix comprised from its arguments.  
For a block matrix $Z$ with $n\in\mathbb{N}$
row partitions and $m\in\mathbb{N}$ column partitions, we denote
by $(Z)_{ij}$ as the matrix block on the $i$th row and $j$th column
in $Z$, where $i\in\underline{n}$ and $j\in\underline{m}$. Denote by
$\text{{\bf row}}_{i}(Z)$ as $[(Z)_{i1},(Z)_{i2},\ldots,(Z)_{in}].$
Let $\text{{\bf gcd}}\left\{ k_{1},k_{2},\cdots,k_{m}\right\} $ signify
the greatest common divisor of a set of integers $k_{1},k_{2},\cdots,k_{m}\in\mathbb{N}$.
The $d\times d$ zero matrix and identity matrix are denoted by $\boldsymbol{0}_{d\times d}$
and $I_{d\times d}$, respectively.

\section{Problem Formulation}

Consider a multi-agent network consisting of $n\in\mathbb{N}$ agents
where the state of an agent $i\in\underline{n}$ is denoted by the vector
$\boldsymbol{x}_{i}(t)\in\mathbb{R}^{d}$ with $d\in\mathbb{N}$.
The state of the multi-agent network is denoted by $\boldsymbol{x}(t)=[\boldsymbol{x}_{1}^{T}(t),\boldsymbol{x}_{2}^{T}(t),\ldots,\boldsymbol{x}_{n}^{T}(t)]^{T}\in\mathbb{R}^{dn}$.
The interaction topology of the network is characterized
by a matrix-weighted graph $\mathcal{G}=(\mathcal{V},\mathcal{E},A)$.
The node and edge sets of $\mathcal{G}$ are denoted by $\mathcal{V}=\left\{ v_{1},v_{2},\ldots,v_{n}\right\} $
and $\mathcal{E}\subseteq\mathcal{V}\times\mathcal{V}$, respectively.
The weight on the edge $(v_{i},v_{j})\in\mathcal{E}$ in $\mathcal{G}$
is a symmetric matrix $A_{ij}\in\mathbb{R}^{d\times d}$ such that
$A_{ij}\succeq0$ or $A_{ij}\succ0$ and $A_{ij}=\boldsymbol{0}_{d\times d}$
if $(v_{i},v_{j})\not\in\mathcal{E}$. Thereby, the matrix-valued
adjacency matrix $A=[A_{ij}]\in\mathbb{R}^{dn\times dn}$ is a block
matrix such that the matrix block located in the $i$th row and 
$j$th column is $A_{ij}$. We shall assume that $A_{ij}=A_{ji}$
for all $v_{i}\not\not=v_{j}\in\mathcal{V}$ and $A_{ii}=\boldsymbol{0}_{d\times d}$
for all $v_{i}\in\mathcal{V}$. The neighbor set of an agent $v_{i}\in\mathcal{V}$
is denoted by $\mathcal{N}_{i}=\left\{ v_{j}\in\mathcal{V}\,|\,(v_{i},v_{j})\in\mathcal{E}\right\} $.
The consensus protocol for each agent in a matrix-weighted network
now assumes the form,
\begin{equation}
\dot{\boldsymbol{x}}_{i}(t)=-\sum_{j\in\mathcal{N}_{i}}A_{ij}(\boldsymbol{x}_{i}(t)-\boldsymbol{x}_{j}(t)),i\in\mathcal{V}.\label{equ:matrix-consensus-protocol}
\end{equation}
Denote by $D=\text{{\bf diag}}\left\{ D_{1},D_{2},\cdots,D_{n}\right\} \in\mathbb{R}^{dn}$
as the matrix-valued degree matrix of $\mathcal{G}$, where
$D_{i}=\sum_{j\in\mathcal{N}_{i}}A_{ij}\in\mathbb{R}^{d\times d}$.
The matrix-valued Laplacian is defined as $L=D-A$. 
Controllability of a networked system examines whether the state of its nodes can be steered from any initial
state to an arbitrary desired state in a finite time by manipulating some of the nodes,
referred to as the leader nodes. Let $\boldsymbol{u}=[\boldsymbol{u}_{1}^{\top},\boldsymbol{u}_{2}^{\top},\ldots,\boldsymbol{u}_{m}^{\top}]^{\top}\in\mathbb{R}^{dm}$
be the control input exerted on the leader nodes, where $\boldsymbol{u}_{j}\in\mathbb{R}^{d}$
and $j\in\underline{m}$. Denote by $B=[B_{il}]\in\mathbb{R}^{dn\times dm}$
as the matrix-weighted input matrix where $B_{il}\in\left\{ \boldsymbol{0}_{d\times d},I_{d\times d}\right\} $.
The set of leaders and ``followers" can now be defined as $\mathcal{V}_{\text{leader}}=\left\{ i\in\mathcal{V}\thinspace|\thinspace B_{il}=I_{d\times d}\right\} $
and $\mathcal{V}_{\text{follower}}=\mathcal{V}\setminus\mathcal{V}_{\text{leader}}$,
respectively. As such, the leader-following multi-agent system on\textbf{
}matrix-weighted networks can be characterized by a linear time-invariant system,
\begin{equation}
\dot{\boldsymbol{x}}(t)=-L\boldsymbol{x}(t)+B\boldsymbol{u}(t).\label{eq:controlled-matrix-consensus-overall}
\end{equation}
Hence, the network \eqref{eq:controlled-matrix-consensus-overall}
is controllable from the leader set $\mathcal{V}_{\text{leader}}$ if
and only if the associated controllability matrix,
\begin{equation}
\mathcal{K}(L,B)=\left[B \; -LB \; L^{2}B \; \cdots \; (-L)^{dn-1}B\right],
\end{equation}
has a full row rank, i.e., $\text{{\bf rank}}(\mathcal{K}(L,B))=dn$.
\begin{defn}
The controllable subspace of the system \eqref{eq:controlled-matrix-consensus-overall}
is defined as the range space of $\mathcal{K}(L,B)$, namely, 
\begin{equation}
\langle L|B\rangle=\text{{\bf img}}(B)+L\text{{\bf img}}(B)+\cdots+L^{dn-1}\text{{\bf img}}(B),
\end{equation}
where the summation is with respect to subspace addition.
\end{defn}
In our subsequent discussion, we provide graph-theoretic lower and upper bounds on the
dimension of the controllable subspace $\langle L|B\rangle$.
\section{Lower Bound on the Dimension of the Controllability Subspace}

In this section, we examine the lower bound on the dimension of $\langle L|B\rangle$;
first, let us introduce the necessary graph-theoretic concepts.
\begin{defn}
For a matrix-weighted network $\mathcal{G}=(\mathcal{V},\mathcal{E},A)$,
a node partition $\pi=\left\{ \mathcal{V}_{1},\mathcal{V}_{2},\ldots,\mathcal{V}_{s}\right\} $
is a collection of subsets $\mathcal{V}_{i}\subset\mathcal{V}$
such that $\mathcal{V}=\mathcal{V}_{1}\cup\mathcal{V}_{2}\cup\cdots\cup\mathcal{V}_{s}$
and $\mathcal{V}_{1}\cap\mathcal{V}_{2}\cap\cdots\cap\mathcal{V}_{s}=\emptyset$,
where $i\in\underline{s}$ and $s\in\mathbb{N}$. The matrix-weighted
characteristic matrix $P(\pi)=\text{[\ensuremath{P_{ij}}(\ensuremath{\pi})]}\in\mathbb{R}^{dn\times ds}$
of a node partition $\pi=\left\{ \mathcal{V}_{1},\mathcal{V}_{2},\ldots,\mathcal{V}_{s}\right\} $
is now defined as,
\[
P_{ij}(\pi)=\begin{cases}
I_{d\times d}, & v_{i}\in\mathcal{V}_{j}\\
\boldsymbol{0}_{d\times d}, & v_{i}\notin\mathcal{V}_{j}
\end{cases}.
\]
\end{defn}
For any $\mathcal{Q}\subseteq\mathcal{V}$, denote by $\delta_{|\mathcal{V}|,\mathcal{Q}}$
as a block matrix with $|\mathcal{V}|$ row partitions and one column
partition such that the $q$th $d\times d$ block in $\delta_{|\mathcal{V}|,\mathcal{Q}}$
is $I_{d\times d}$, and all the remaining blocks are $\boldsymbol{0}_{d\times d}$,
where $v_{q}\in\mathcal{Q}$.
\begin{example}
Consider a $5$-node matrix-weighted network with a node partition
$\pi=\left\{ \{1\},\{2,3\},\{4,5\}\right\} $ and the dimension of weight
matrices on edges is $d=2$. Then
\[
P(\pi)=[\delta_{5,\left\{ 1\right\} },\thinspace\delta_{5,\left\{ 2,3\right\} },\thinspace\delta_{5,\left\{ 4,5\right\} }].
\]
\end{example}
A path in a matrix-weighted network $\mathcal{G}=(\mathcal{V},\mathcal{E},A)$
is a sequence of edges of the form $$\mathcal{P}_{v_{i_{1}},v_{i_{p}}}=(v_{i_{1}},v_{i_{2}}),(v_{i_{2}},v_{i_{3}}),\ldots,(v_{i_{p-1}},v_{i_{p}}),$$
where nodes $v_{i_{1}},v_{i_{2}},\ldots,v_{i_{p}}\in\mathcal{V}$
are all distinct and it is said that $v_{i_{p}}$ is reachable from
$v_{i_{1}}$; a path $\mathcal{P}_{v_{i_{1}},v_{i_{p}}}$ turns to a cycle
if $v_{i_{1}}=v_{i_{p}}$. The network $\mathcal{G}$ is connected
if any two distinct nodes in $\mathcal{G}$ are reachable from each
other. A tree is a connected graph with $n$ nodes and $n-1$ edges
where $n\in\mathbb{N}$. All networks discussed in this paper are assumed to 
be connected. The shortest path between two nodes $v_{i},v_{j}\in\mathcal{V}$
is a path that contains the least number of the edges; the
number of the edges on this shortest path is referred to as the distance
between nodes $v_{i}$ and $v_{j}$, denoted by $\text{{\bf dist}}(v_{i},v_{j})$.
The diameter of $\mathcal{G}$ is then defined as,
\[
\text{{\bf diam}}(\mathcal{G})={\displaystyle \max_{v_{i},v_{j}\in\mathcal{V}}}\text{{\bf dist}}(v_{i},v_{j}).
\]
An edge $(v_{i},v_{j})\in\mathcal{E}$ is positive definite or positive
semi-definite if its weight matrix $A_{ij}$ is positive definite
or positive semi-definite. 
\begin{defn}[Positive definite path]
 A positive definite path in a matrix-weighted network
$\mathcal{G}$ is a path for which every edge has a positive definite weight. 
\end{defn}
In the subsequent discussion, we will characterize a lower bound on the dimension
of the controllable subspace of \eqref{eq:controlled-matrix-consensus-overall}
for acyclic networks, followed by cycle and complete networks.
%
In particular, we examine the influence of the
positive definiteness of weight matrices on $\text{{\bf dim}}(\langle L|B\rangle)$. 
\begin{defn}[Distance partition]
\label{def: distance-partition} Let $\mathcal{G}=(\mathcal{V},\mathcal{E},A)$
be a matrix-weighted network. The distance partition relative to an
agent $v_{i}\in\mathcal{V}$ consists of the subsets,
\[
\mathcal{C}_{r}=\left\{ v_{j}\in\mathcal{V}\thinspace|\thinspace\text{{\bf dist}}(v_{i},v_{j})=r\right\},
\]
where $0\leq r\leq\text{{\bf diam}}(\mathcal{G})$.
\end{defn}
\begin{thm}
\label{thm:2}Let $\mathcal{G}=(\mathcal{V},\mathcal{E},A)$ be a
matrix-weighted tree network whose dimension of the weight matrix
is $d\in\mathbb{N}$. Let $v_{l}\in\mathcal{V}$ be the leader agent
and denote the distance partition relative to $v_{l}$ as $\pi_{D}(v_{l})=\left\{ \mathcal{C}_{0},\mathcal{C}_{1},\ldots,\mathcal{C}_{r}\right\}$,
where $0\leq r\leq\text{{\bf diam}}(\mathcal{G})$. If there exists
an agent $v_{i}$ in $\mathcal{C}_{r}$ such that the path $\mathcal{P}_{v_{l},v_{i}}$
is  positive definite, then \textup{$\text{{\bf dim}}(\langle L|B\rangle)\geq d\thinspace|\pi_{D}(v_{l})|$.}
\end{thm}
\begin{IEEEproof}
The adopted line of reasoning is similar to that presented in~\cite{zhang2011controllability} for the scalar-weights.
Without loss of generality, let $v_{1}$ be the leader agent. Denote
the distance partition relative to $v_{1}$ as $\pi_{D}(v_{1})=\left\{ \mathcal{C}_{0},\mathcal{C}_{1},\ldots,\mathcal{C}_{r}\right\} $.
Specifically, 
\begin{align*}
\mathcal{C}_{0} & =v_{1},\\
\mathcal{C}_{q} & =\left\{ v_{i}\thinspace|\thinspace v_{i}\in\mathcal{V},\text{{\bf dist}}(v_{i},v_{1})=q\right\} \\
 & =\left\{ v_{1}^{(q)},v_{2}^{(q)},\cdots,v_{|\mathcal{C}_{q}|}^{(q)}\right\} ,q\in\underline{r}.
\end{align*}

According to Definition \ref{def: distance-partition}, there
does not exist agents in $\mathcal{C}_{i}$ with a neighbor in $\mathcal{C}_{j}$
if $|i-j|>1$, where $i,j\in\underline{r}$. Then the matrix-weighted
Laplacian of $\mathcal{G}$ admits the form,
\[
L=\left[\begin{array}{cccccc}
L_{00} & L_{01} & \boldsymbol{0} & \cdots & \boldsymbol{0} & \boldsymbol{0}\\
L_{10} & L_{11} & L_{12} & \cdots & \boldsymbol{0} & \boldsymbol{0}\\
\boldsymbol{0} & L_{21} & L_{22} & \cdots & \boldsymbol{0} & \boldsymbol{0}\\
\vdots & \vdots & \vdots & \ddots & \vdots & \vdots\\
\boldsymbol{0} & \boldsymbol{0} & \boldsymbol{0} & \cdots & L_{r-1,r-1} & L_{r-1,r}\\
\boldsymbol{0} & \boldsymbol{0} & \boldsymbol{0} & \cdots & L_{r,r-1} & L_{r,r}
\end{array}\right],
\]
where $L_{kl}\in\mathbb{R}^{d\thinspace|\mathcal{C}_{k}|\times d\thinspace|\mathcal{C}_{l}|}$
for all $0\leq k,l\leq r$ and $\boldsymbol{0}$'s are zero matrices with
proper dimensions.

Let $E=[\begin{array}{cccc}
B & LB & \cdots & L^{r}B\end{array}]$ be a block matrix with $r+1$ row partitions and $r+1$ column partitions.
Note that as agent $v_{1}$ is the leader, $B=[\begin{array}{cccc}
I_{d\times d} & 0_{d\times d} & \cdots & 0_{d\times d}\end{array}]^{\top}$ and,
\begin{eqnarray*}
E & = & \left[\begin{array}{cccc}
E_{00} & E_{01} & \cdots & E_{0,r}\\
E_{10} & E_{11} & \cdots & E_{1,r}\\
\vdots & \vdots & \ddots & \vdots\\
E_{r,0} & E_{r,1} & \cdots & E_{r,r}
\end{array}\right],
\end{eqnarray*}
where $E_{00}=I_{d\times d}$, $E_{qq}$ is a block matrix with $|\mathcal{C}_{q}|$
row partitions and $1$ column partition where $q\in\underline{r}$
and $E_{pq}$ with $p>q$ are matrices with proper size and all elements
equal to $0$ where $p\in\underline{r}$.
In particular, we are interested in those blocks located in $q$th
row and $q$th column in $E$ since they are crucial in determining
$\text{{\bf rank}}(E)$.

Denote the block in $s$th row block in $E_{qq}$ as $E_{qq}^{(s)}$,
where $s\in\left\{ 1,2,\cdots,|\mathcal{C}_{q}|\right\} $ and $q\in\underline{r}$;
then 
\[
E_{qq}^{(s)}=\underset{(i,j)\in\ensuremath{\mathcal{P}}_{v_{1},v_{s}^{(q)}}}{\prod}A_{ij}.
\]

By our standing assumption, there exists one node in $\mathcal{C}_{r}$ such that the path between
this node and $v_{l}$ are positive definite. As the product
of positive definite matrices has full rank,
one has $\text{{\bf rank}}(E_{qq})=d$ for all $q\in\underline{r}$.
Hence, 
\[
\text{{\bf rank}}(E)=d\thinspace|\pi_{D}(v_{1})|,
\]
completing the proof.
\end{IEEEproof}
\begin{cor}[Path network]
 Let $\mathcal{G}=(\mathcal{V},\mathcal{E},A)$ be a path network
in the form of $$\mathcal{P}_{v_{i_{1}},v_{i_{p}}}=(v_{i_{1}},v_{i_{2}}),(v_{i_{2}},v_{i_{3}}),\ldots,(v_{i_{p-1}},v_{i_{p}}),$$
where nodes $v_{i_{1}},v_{i_{2}},\ldots,v_{i_{p}}\in\mathcal{V}$.
Then $\mathcal{G}$ is controllable from $v_{i_{1}}$ (or $v_{i_{p}}$)
if and only if the path $\mathcal{P}_{v_{i_{1}},v_{i_{p}}}$ is positive
definite.
\end{cor}
\begin{cor}[Cycle network]
Let $\mathcal{G}=(\mathcal{V},\mathcal{E},A)$
be a matrix-weighted cycle with dimension of the weight
matrix as $d\in\mathbb{N}$. Let $v_{l}\in\mathcal{V}$ be a leader
agent and denote the distance partition relative to $v_{l}$ as $\pi_{D}(v_{l})=\left\{ \mathcal{C}_{0},\mathcal{C}_{1},\ldots,\mathcal{C}_{r}\right\}$,
where,
\[
r=\begin{cases}
\frac{|\mathcal{V}|}{2}, & |\mathcal{V}|\thinspace\text{is even};\\
\frac{|\mathcal{V}|-1}{2}, & |\mathcal{V}|\thinspace\text{is odd}.
\end{cases}
\]
If there exists an agent $v_{i}$ in $\mathcal{C}_{r}$ such that
the shortest path between $v_{i}$ and $v_{l}$ are positive definite,
then \textup{
\[
\text{{\bf dim}}(\langle L|B\rangle)\geq\begin{cases}
d\thinspace\frac{|\mathcal{V}|}{2}+1, & |\mathcal{V}|\thinspace\text{is even};\\
d\thinspace\frac{|\mathcal{V}|+1}{2}, & |\mathcal{V}|\thinspace\text{is odd}.
\end{cases}
\]
}
\end{cor}
\begin{cor}[Complete network]
Let $\mathcal{G}=(\mathcal{V},\mathcal{E},A)$
be a matrix-weighted complete network with the dimension of the weight
matrix as $d\in\mathbb{N}$. Let $v_{l}\in\mathcal{V}$ be a leader
agent and denote the distance partition relative to $v_{l}$ as $\pi_{D}(v_{l})=\left\{ \mathcal{C}_{0},\mathcal{C}_{1}\right\} $.
If there exists an agent $v_{i}$ in $\mathcal{C}_{1}$ such that
the path $\mathcal{P}_{v_{l},v_{i}}$ is positive definite, then \textup{$\text{{\bf dim}}(\langle L|B\rangle)\geq d$.}
\end{cor}
Note that from Theorem \ref{thm:2}, the rank of weight
matrices influences the lower bound on the dimension of the
controllable subspace of \eqref{eq:controlled-matrix-consensus-overall};
this is distinct from the scalar-weighted case.
As such, the semi-definiteness of weight matrices plays an important 
role in the controllability of matrix weighted networks.

\section{Upper Bound on the Dimension of the Controllability Subspace}

 We now proceed to examine graph-theoretic characterizations of the upper
bound of the controllable subspace of system \eqref{eq:controlled-matrix-consensus-overall}
in terms of the almost equitable partition. For a given subset $\mathcal{Q}\in\mathcal{V}$
in a matrix-weighted network $\mathcal{G}=(\mathcal{V},\mathcal{E},A)$
and an agent $v_{i}\in\mathcal{V}$, denote the matrix-valued degree
of $v_{i}$ relative to $\mathcal{Q}$ as,
\[
D(v_{i},\mathcal{Q})=\sum_{v_{j}\in\mathcal{Q}}A_{ij}.
\]

\begin{defn}
\label{def:AEP}An $s-$partition $\pi=\left\{ \mathcal{V}_{1},\mathcal{V}_{2},\ldots,\mathcal{V}_{s}\right\} $
of a matrix-weighted network $\mathcal{G}=(\mathcal{V},\mathcal{E},A)$
is an almost equitable partition if for $\forall i\not=j\in\underline{s}$
and $\forall v,w\in\mathcal{V}_{i}$ one has $D(v,\mathcal{V}_{j})=D(w,\mathcal{V}_{j})$.
\end{defn}
According to Definition \ref{def:AEP}, if an $s-$partition, $\pi=\left\{ \mathcal{V}_{1},\mathcal{V}_{2},\ldots,\mathcal{V}_{s}\right\} $
is an almost equitable partition, then one can denote $D(\mathcal{V}_{i},\mathcal{V}_{j})=D(v,\mathcal{V}_{j})$
for $\forall v\in\mathcal{V}_{i}$. Next, we proceed to define the
quotient graph of a matrix-weighted network based on the almost equitable
partition.
\begin{defn}
For a given almost equitable partition $\pi=\left\{ \mathcal{V}_{1},\mathcal{V}_{2},\ldots,\mathcal{V}_{s}\right\} $
of a matrix-weighted network $\mathcal{G}=(\mathcal{V},\mathcal{E},A)$,
the quotient graph of $\mathcal{G}$ over $\pi$ is a matrix-weighted
network denoted by $\mathcal{G}/\pi$ with the node set,
\[
\mathcal{V}(\mathcal{G}/\pi)=\left\{ \mathcal{V}_{1},\mathcal{V}_{2},\ldots,\mathcal{V}_{s}\right\} ,
\]
whose edge set is,
\[
\mathcal{E}(\mathcal{G}/\pi)=\left\{ (\mathcal{V}_{i},\mathcal{V}_{j})\thinspace|\thinspace D(\mathcal{V}_{i},\mathcal{V}_{j})\neq\boldsymbol{0}_{d\times d}\right\} ,
\]
and the weight on edge $(\mathcal{V}_{i},\mathcal{V}_{j})$ is $D(\mathcal{V}_{i},\mathcal{V}_{j})$
for $i\not=j\in\underline{s}$.
\end{defn}
Note that the condition $D(\mathcal{V}_{i},\mathcal{V}_{j})=D(\mathcal{V}_{j},\mathcal{V}_{i})$
does not necessary hold; as such, the quotient graph $\mathcal{G}/\pi$
can be directed. The following result provides the relationship between
the $L-$invariant subspace and the almost equitable partition of
matrix-weighted networks. 
\begin{lem}
\label{thm:L-invariant theorem}Let $\mathcal{G}=(\mathcal{V},\mathcal{E},A)$
be a matrix-weighted network with the dimension of edge weight $d\in\mathbb{N}$,
$L$ be the matrix-valued Laplacian of $\mathcal{G}$, $\pi=\left\{ \mathcal{V}_{1},\mathcal{V}_{2},\ldots,\mathcal{V}_{s}\right\} $
be a $s-$partition of $\mathcal{V}(\mathcal{G})$ and $P(\pi)$ be
the characteristic matrix of $\pi$. Then $\pi$ is an almost equitable
partition of $\mathcal{G}$ if and only if $\text{{\bf img}}(P(\pi))$
is $L-$invariant, i.e., there exists a matrix $L^{\pi}\in\mathbb{R}^{ds\times ds}$
such that $LP(\pi)=P(\pi)L^{\pi}.$
\end{lem}
\begin{IEEEproof}
(Necessity) Define the matrix $L^{\pi}\in\mathbb{R}^{ds\times ds}$
as
\[
(L^{\pi})_{ij}=\begin{cases}
{\displaystyle \sum_{\mathcal{V}_{j}\in\mathcal{V}(\mathcal{G}/\pi)}}D(\mathcal{V}_{i},\mathcal{V}_{j}), & i=j;\\
-D(\mathcal{V}_{i},\mathcal{V}_{j}), & i\neq j.
\end{cases}
\]
Suppose that $\pi=\left\{ \mathcal{V}_{1},\mathcal{V}_{2},\ldots,\mathcal{V}_{s}\right\} $
is an almost equitable partition of the matrix-weighted network $\mathcal{G}$
and $v_{p}\in\mathcal{V}_{k}$, where $p\in\underline{n}$ and $k\in\underline{s}$.
On one hand, the $p$-th block row of $LP(\pi)$ can be
characterized by,
\begin{align*}
(LP(\pi))_{p} & =[-\sum_{j\in\mathcal{V}_{1}\cap\mathcal{N}_{\mathcal{G}}(v_{p})}A_{pj},-\sum_{j\in\mathcal{V}_{2}\cap\mathcal{N}_{\mathcal{G}}(v_{p})}A_{pj},\ldots,\\
 & D_{p}-\sum_{j\in\mathcal{V}_{k}\cap N_{\mathcal{G}}(v_{p})}A_{pj},-\sum_{j\in\mathcal{V}_{k+1}\cap N_{\mathcal{G}}(v_{p})}A_{pj},\\
 & \ldots,-\sum_{j\in\mathcal{V}_{s}\cap N_{\mathcal{G}}(v_{p})}A_{pj}].
\end{align*}
On the other hand, the entries in the $p$-th block row of $P(\pi)L^{\pi}$
are,
\begin{align*}
(P(\pi)L^{\pi})_{p} & =[-D(\mathcal{V}_{k},\mathcal{V}_{1}),\ldots,-D(\mathcal{V}_{k},\mathcal{V}_{k-1}),\sum_{r\neq k}D(\mathcal{V}_{k},\mathcal{V}_{r}),\\
 & -D(\mathcal{V}_{k},\mathcal{V}_{k+1}),\ldots,-D(\mathcal{V}_{k},\mathcal{V}_{s})].
\end{align*}
According to Definition \ref{def:AEP}, we have
\[
\sum_{j\in\mathcal{V}_{r}\cap\mathcal{N}_{\mathcal{G}}(p)}A_{pj}=D(\mathcal{V}_{k},\mathcal{V}_{r}),
\]
and 
\[
D_{p}-\sum_{j\in\mathcal{V}_{k}\cap N_{\mathcal{G}}(p)}A_{pj}=\sum_{r\neq k}D(\mathcal{V}_{k},\mathcal{V}_{r}).
\]
Then 
\[
\text{{\bf row}}_{p}(LP(\pi))=\text{{\bf row}}_{p}(P(\pi)L^{\pi}),
\]
which implies that $LP(\pi)=P(\pi)L^{\pi}.$

(Sufficiency) Suppose that $\pi$ is an $s-$partition of the matrix-weighted
network $\mathcal{G}$ satisfying $LP(\pi)=P(\pi)L^{\pi}$. Then each
column in $LP(\pi)$ is the linear combination of the columns in $P(\pi)$.
For each block column of $LP(\pi)$, the matrix blocks corresponding
to the agents belonging to the same subset in $\pi$ are identical.
Therefore one has,
\[
(LP(\pi))_{ij}=-\sum_{r\in\mathcal{V}_{j}\cap N_{\mathcal{G}}(i)}A_{ir},\forall i\neq j,
\]
and for any $k$ in the same subset as $i$, 
\[
(LP(\pi))_{kj}=-\sum_{r\in\mathcal{V}_{j}\cap N_{\mathcal{G}}(k)}A_{kr}.
\]
Note that $(LP(\pi))_{ij}=(LP(\pi))_{kj}$ implies that,
\[
\sum_{r\in\mathcal{V}_{j}\cap N_{\mathcal{G}}(i)}A_{ir}=\sum_{r\in\mathcal{V}_{j}\cap N_{\mathcal{G}}(k)}A_{kr},
\]
for any $k$ in the same subset as $i$. Therefore, $\pi$ is an almost
equitable partition.
\end{IEEEproof}
Lemma \ref{thm:L-invariant theorem} has the following immediate consequence.
\begin{thm}
\label{thm:the upper bound}Let $\mathcal{G}=(\mathcal{V},\mathcal{E},A)$
be a matrix-weighted network with the dimension of edge weight $d\in\mathbb{N}$.
Suppose that $\pi=\left\{ \mathcal{V}_{1},\mathcal{V}_{2},\ldots,\mathcal{V}_{s}\right\} $
is an almost equitable partition of $\mathcal{G}$ with the characteristic
matrix $P(\pi)$ where $1\leq s<n$. Denote $B=[b_{1},b_{2},\cdots,b_{m}]\in\mathbb{R}^{dn\times dm}$
as the input matrix where $b_{i}\in\left\{ \boldsymbol{0}_{d\times d},I_{d\times d}\right\} ^{n}$,
the matrix blocks in $b_{i}$ corresponding to the agents belonging
to the same subset in $\pi$ are the same and $i\in\underline{m}$.
Then, (1) $\langle L|B\rangle\subseteq\text{{\bf img}}(P(\pi))$,
(2) $\text{{\bf dim}}(\langle L|B\rangle)\leq ds$, and (3) the pair $(L,B)$
is uncontrollable.
\end{thm}
\begin{IEEEproof}
Since the matrix blocks in $b_{i}$ corresponding to the agents belonging
to the same subset in $\pi$ are the same where $i\in\underline{m}$,
then $\text{{\bf img}}(B)\subseteq\text{{\bf img}}(P(\pi))$. In the
meantime, $\text{{\bf img}}(P(\pi))$ is $L-$invariant according
to Lemma \ref{thm:L-invariant theorem}; thus we have,
\begin{align*}
\langle L|B\rangle & =\text{{\bf img}}(B)+L\text{{\bf img}}(B)+\cdots+L^{dn-1}\text{{\bf img}}(B)\\
 & \subseteq\text{{\bf img}}(P(\pi))+L\text{{\bf img}}(P(\pi))\cdots+L^{dn-1}\text{{\bf img}}(P(\pi))\\
 & =\text{{\bf img}}(P(\pi)),
\end{align*}
implying $\text{{\bf dim}}(\langle L|B\rangle)\leq ds$. Since $1\leq ds<dn$,
the pair $(L,B)$ is uncontrollable.
\end{IEEEproof}

\section{On Uncontrollable Input Matrix}

Note from that Theorem \ref{thm:the upper bound} provides an upper
bound on the controllable subspace using the range space of the characteristic
matrix of the almost equitable partition. It is shown that $\text{{\bf img}}(B)\subseteq\text{{\bf img}}(P(\pi))$
can directly lead to the uncontrollability of the network when the
almost equitable partition $\pi=\left\{ \mathcal{V}_{1},\mathcal{V}_{2},\ldots,\mathcal{V}_{s}\right\} $
is non-trivial. However, is there any other leader selections that
induces the uncontrollability of $(L,B)$? In the following discussions,
we proceed to provide the structure of the uncontrollable matrix $B$.
\begin{thm}
\label{thm:choose of uncontrollable b}Let $\mathcal{G}=(\mathcal{V},\mathcal{E},A)$
be a matrix-weighted network with the dimension of edge weight $d\in\mathbb{N}$.
Suppose that $\pi=\left\{ \mathcal{V}_{1},\mathcal{V}_{2},\ldots,\mathcal{V}_{s}\right\} $
is an almost equitable partition of $\mathcal{G}$ with the characteristic
matrix $P(\pi)=[P_{1},P_{2},\cdots,P_{s}]$, where $P_{1},P_{2},\ldots,P_{s}\in\mathbb{R}^{dn\times d}$
and $1\leq s<n$. Let $\pi$ be reducible and $q_{j}=\frac{|\mathcal{V}_{j}|}{{\bf gcd}(\pi)}$
where $j=\underline{s}.$ Let $B\in\left\{ 0_{d\times d},I_{d\times d}\right\} ^{n}$
be such that 
\begin{equation}
\boldsymbol{p}_{j1}^{\top}\boldsymbol{b}_{1}=\boldsymbol{p}_{j2}^{\top}\boldsymbol{b}_{2}=\cdots=\boldsymbol{p}_{jd}^{\top}\boldsymbol{b}_{d}=cq_{j},\label{eq:the condition}
\end{equation}
where $c$ is an integer such that $1\leq c\leq{\bf gcd}(\pi)-1$,
$\boldsymbol{p}_{j1},\boldsymbol{p}_{j2},\ldots,\boldsymbol{p}_{jd}\in\mathbb{R}^{dn\times1}$
and $\boldsymbol{b}_{1,}\boldsymbol{b}_{2},\ldots,\boldsymbol{b}_{d}\in\mathbb{R}^{dn\times1}$
are columns of matrices $P_{j}$ and $B$, respectively. Then $(L,B)$
is uncontrollable.
\end{thm}
\begin{IEEEproof}
Since $\text{{\bf img}}(P(\pi))$ is $L-$invariant,
there exists an eigenvector $\boldsymbol{w}\in\text{{\bf img}}(P(\pi))$
of $L$ satisfying $\boldsymbol{w}\notin\text{{\bf span}}\{\boldsymbol{1}_{n}\otimes I_{d}\}$
and $\boldsymbol{w}^{\top}\boldsymbol{1}_{dn}=0.$ Note that
$\{\boldsymbol{p}_{11},\cdots,\boldsymbol{p}_{1d},\boldsymbol{p}_{21},\cdots,\boldsymbol{p}_{2d},\cdots,\boldsymbol{p}_{s1},\cdots,\boldsymbol{p}_{sd}\}$
forms a basis of $\text{{\bf img}}(P(\pi))$; as such, 
\[
\boldsymbol{w}=\sum_{j=1}^{s}\sum_{t=1}^{d}\alpha_{jt}\boldsymbol{p}_{jt}
\]
for some $\alpha_{jt}$'s.
Due to the fact that 
\begin{align*}
\boldsymbol{w}^{\top}(\boldsymbol{1}_{n}\otimes I_{d}) & =\left(\sum_{j=1}^{s}\alpha_{j1}|\mathcal{V}_{j}|,\cdots,\sum_{j=1}^{s}\alpha_{jd}|\mathcal{V}_{j}|\right)\\
 & =[0,\cdots,0]\in\mathbb{R}^{1\times dn},
\end{align*}
if we choose $B=[\boldsymbol{b}_{1},\cdots,\boldsymbol{b}_{d}]$
satisfying \eqref{eq:the condition}, then we have 
\begin{align*}
\boldsymbol{w}^{\top}\boldsymbol{b}_{k} & =\left(\sum_{j=1}^{s}\sum_{t=1}^{d}\alpha_{jt}\boldsymbol{p}_{jt}^{\top}\right)\boldsymbol{b}_{k}=\left(\sum_{j=1}^{s}\alpha_{jk}\boldsymbol{p}_{jk}^{\top}\right)\boldsymbol{b}_{k}\\
 & =\sum_{j=1}^{s}\alpha_{jk}cq_{j}=\sum_{j=1}^{s}\alpha_{jk}c\frac{|\mathcal{V}_{j}|}{\text{{\bf gcd}}(\pi)}\\
 & =\frac{c}{\text{{\bf gcd}}(\pi)}\sum_{j=1}^{s}\alpha_{jk}|\mathcal{V}_{j}|=0,
\end{align*}
for any $k\in\underline{d}$. Therefore, $(L,B)$ is uncontrollable.
\end{IEEEproof}

\section{Examples}

\begin{figure}[tbh]
\begin{centering}
\begin{tikzpicture}[scale=0.8]
	
	\node (n1) at (-1,1) [circle,fill=black!20,draw] {1};
    \node (n2) at (1,1) [circle,fill=black!20,draw] {2};
    \node (n3) at (2,0) [circle,fill=black!20,draw] {3};
    \node (n4) at (1,-1) [circle,fill=black!20,draw] {4};
    \node (n5) at (-1,-1) [circle,fill=black!20,draw] {5};
    \node (n6) at (-2,0) [circle,fill=black!20,draw] {6};



	\draw[-, thick, color=black!50] (n1) -- (n2); 
	\draw[-, thick, color=black!50] (n2) -- (n3); 
	\draw[-, thick, color=black!50] (n3) -- (n4);
    \draw[-, thick, color=black!50] (n4) -- (n5);
    \draw[-, thick, color=black!50] (n1) -- (n6); 
    \draw[-, thick, color=black!50] (n1) -- (n4); 
    \draw[-, thick, color=black!50] (n2) -- (n5); 
    \draw[-, thick, color=black!50] (n5) -- (n6); 

\end{tikzpicture}
\par\end{centering}
\caption{\textcolor{black}{A matrix-weighted network with $5$ nodes.}}
\label{fig:example-network-2}
\end{figure}
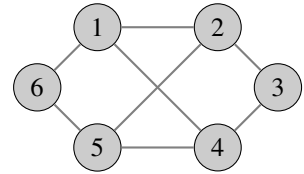

We now provide examples to demonstrate
the results discussed in the paper. The first example
underscores that the semi-definiteness of edge weights can have an adverse effect on the controllability of 
matrix-weighted networks; this was examined in Theorem \ref{thm:2}.
\begin{example}
\label{exa:1}Consider the matrix-weighted path network in Figure
\ref{fig:example-network}. Choose agent $1$ as the leader and set the weight
matrices on edges as,
\[
A_{12}=\left[\begin{array}{cc}
1 & 1\\
1 & 2
\end{array}\right],A_{23}=\left[\begin{array}{cc}
1 & 0\\
0 & 2
\end{array}\right],
\]
\[
A_{34}=\left[\begin{array}{cc}
2 & 1\\
1 & 2
\end{array}\right],A_{45}=\left[\begin{array}{cc}
1 & 2\\
2 & 5
\end{array}\right];
\]
note that the weight matrices are all positive definite.
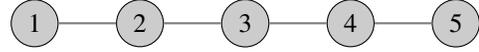
\begin{figure}[tbh]
\begin{centering}
\begin{tikzpicture}[scale=0.7]
	
	\node (n1) at (-4,0) [circle,fill=black!20,draw] {1};
    \node (n2) at (-2,0) [circle,fill=black!20,draw] {2};
    \node (n3) at (0,0) [circle,fill=black!20,draw] {3};
    \node (n4) at (2,0) [circle,fill=black!20,draw] {4};
    \node (n5) at (4,0) [circle,fill=black!20,draw] {5};



	\draw[-, thick, color=black!50] (n1) -- (n2); 
	\draw[-, thick, color=black!50] (n2) -- (n3); 
	\draw[-, thick, color=black!50] (n3) -- (n4);
    \draw[-, thick, color=black!50] (n4) -- (n5);

\end{tikzpicture}
\par\end{centering}
\caption{\textcolor{black}{A matrix-weighted path }network\textcolor{black}{{}
with $5$ nodes.}}
\label{fig:example-network}
\end{figure}

The matrix-valued Laplacian $L$ and the matrix-weighted input
matrix can now be written as, \begin{footnotesize} 
\[
L=\left[\begin{array}{cccccccccc}
1 & 1 & -1 & -1 & 0 & 0 & 0 & 0 & 0 & 0\\
1 & 2 & -1 & -2 & 0 & 0 & 0 & 0 & 0 & 0\\
-1 & -1 & 2 & 1 & -1 & 0 & 0 & 0 & 0 & 0\\
-1 & -2 & 1 & 4 & 0 & -2 & 0 & 0 & 0 & 0\\
0 & 0 & -1 & 0 & 3 & 1 & -2 & -1 & 0 & 0\\
0 & 0 & 0 & -2 & 1 & 4 & -1 & -2 & 0 & 0\\
0 & 0 & 0 & 0 & -2 & -1 & 3 & 3 & -1 & -2\\
0 & 0 & 0 & 0 & -1 & -2 & 3 & 7 & -2 & -5\\
0 & 0 & 0 & 0 & 0 & 0 & -1 & -2 & 1 & 2\\
0 & 0 & 0 & 0 & 0 & 0 & -2 & -5 & 2 & 5
\end{array}\right],
\]
\end{footnotesize}and $B=[\delta_{10,\left\{ 1\right\} },\thinspace\delta_{10,\left\{ 2\right\} }].$
The dimension of the controllable subspace $\langle L|B\rangle$
in this example is $10$ and therefore $(L,B)$ is controllable. We
proceed to replace the weight matrix between agent $2$ and agent
$3$ by a positive semi-definite matrix
\[
A_{23}=\left[\begin{array}{cc}
1 & 1\\
1 & 1
\end{array}\right].
\]
In this case, the dimension of the controllable subspace $\langle L|B\rangle$
becomes $9$, implying that $(L,B)$ is uncontrollable.
\end{example}
The next example illustrates the results presented 
in the Theorem \ref{thm:choose of uncontrollable b}.
\begin{example}
\label{exa:2}Consider the matrix-weighted network in Figure \ref{fig:example-network-2}.
The weight matrices are chosen as,
\[
A_{16}=A_{25}=A_{14}=A_{23}=\left[\begin{array}{cc}
1 & 1\\
1 & 2
\end{array}\right],
\]
\end{example}
\[
A_{12}=A_{34}=A_{45}=A_{56}=\left[\begin{array}{cc}
1 & 1\\
1 & 1
\end{array}\right].
\]

Note that the network in Figure \ref{fig:example-network-2}
has an almost equitable partition $\pi=\left\{ \mathcal{V}_{1},\mathcal{V}_{2}\right\}$,
where $\mathcal{V}_{1}=\{1,2\}$ and $\mathcal{V}_{2}=\{3,4,5,6\}$.
The characteristic matrix $P(\pi)=[P_{1},P_{2}]$ has entries $P_{1}=[\delta_{6,\left\{ 1,2\right\} }]$
and $P_{2}=[\delta_{6,\left\{ 3,4,5,6\right\} }]$. Since $\text{{\bf gcd}}(\pi)=2$,
$q_{1}=1$, $q_{2}=2$. Choose the input matrix as,
\[
B=[\delta_{6,\left\{ 1,3,6\right\} }],
\]
that satisfies $\boldsymbol{p}_{11}^{\top}\boldsymbol{b}_{1}=\boldsymbol{p}_{12}^{\top}\boldsymbol{b}_{2}=1$
and $\boldsymbol{p}_{21}^{\top}\boldsymbol{b}_{1}=\boldsymbol{p}_{22}^{\top}\boldsymbol{b}_{2}=2$.
Then $\text{{\bf rank}}(\mathcal{K}(L,B))=9$, implying that the $(L,B)$
in this example is uncontrollable, which is consistent with Theorem \ref{thm:choose of uncontrollable b}.

\section{Conclusion}

This paper examines the controllability problem of multi-agent system
on matrix-weighed networks. Both lower and upper bounds on
the dimension of the controllable subspace--associated with controlled
consensus dynamics on matrix-weighted networks--is provided from a
graph-theoretic perspective. The structure of an uncontrollable input
matrix is further investigated. Examples are then provided to demonstrate
the theoretical results.

In our further work, we will examine the
graph-theoretic characterizations of lower/upper bound of controllable
subspace of matrix-weighted networks allowing both positive (semi-)definite
and negative (semi-)definite weight matrices.

\bibliographystyle{IEEEtran}

\bibliography{mybib}

\end{document}